# Switching of magnetic ground states across the UIr$_{1-x}$Rh$_x$Ge alloy system

Jiří Pospíšil[1,2*], Yoshinori Haga[1], Shinsaku Kambe[1], Yo Tokunaga[1], Naoyuki Tateiwa[1], Dai Aoki[3], Fuminori Honda[3], Ai Nakamura[3], Yoshiya Homma[3], Etsuji Yamamoto[1] and Tomoo Yamamura[4]

[1] *Advanced Science Research Center, Japan Atomic Energy Agency, Tokai, Ibaraki, 319-1195, Japan*

[2] *Charles University in Prague, Faculty of Mathematics and Physics, Department of Condensed Matter Physics, Ke Karlovu 5, 121 16 Prague 2, Czechia*

[3] *Institute for Materials Research, Tohoku University, Oarai, Ibaraki 311-1313, Japan*

[4] *Institute for Materials Research, Tohoku University, 2-1-1, Katahira, Aoba, Sendai, Miyagi 980-8577, Japan*

## Abstract

We investigated the evolution of magnetism in the UIr$_{1-x}$Rh$_x$Ge system by the systematic study of high-quality single crystals. Lattice parameters of both parent compounds are very similar resulting in almost identical nearest interatomic uranium distance close to the Hill limit. We established the $x$-$T$ phase diagram of the UIr$_{1-x}$Rh$_x$Ge system and found a discontinuous antiferromagnetic/ferromagnetic boundary at $x_{\mathrm{crit}} = 0.56$ where a local minimum in ordering temperature and maximum of the Sommerfeld coefficient $\gamma \approx 175$ mJ/mol K$^2$ occurs in the UCoGe-URhGe-UIrGe system, signaling an increase in magnetic fluctuations. However, a quantum critical point is not realized because of the finite ordering temperature at $x_{\mathrm{crit}}$. A magnon gap on the antiferromagnetic side abruptly suppresses magnetic fluctuations. We find a field-induced first order transition in the vicinity of the critical magnetic field along the $b$ axis in the entire UIr$_{1-x}$Rh$_x$Ge system including the ferromagnetic region ~UCo$_{0.6}$Rh$_{0.4}$Ge - URhGe.

*E-mail address: jiri.pospisil@centrum.cz

## I. Introduction

Uranium intermetallics with 5$f$ electrons at the boundary between localized and itinerant character are of continuing interest. The crossover was empirically established by Hill[1] at an interatomic uranium-uranium distance $d_{\mathrm{U-U}} \approx 3.5$ Å. Exotic electronic phenomena often appear in compounds satisfying this criterion. Ferromagnetic superconductors (FM SC) URhGe[2] and UCoGe[3] are exemplary cases. Recent papers on the related isostructural TiNiSi-type U$T$Ge ($T$ = transition metal) compounds found their magnetism scaling according to the Hill criterion[4,5] and uncovered another promising candidate, UIrGe.

UIrGe[6] has an almost identical nearest interatomic uranium-uranium distance to URhGe but orders antiferromagnetically[7] with Néel temperature $T_{\mathrm{N}} = 16.5$ K. The magnetic structure of UIrGe consists of FM zig-zag chains along the $a$ axis[8,9] resembling the magnetic structures of the FMs UCoGe[10-12] and URhGe. The chains are antiferromagnetically coupled. A spin-flop transition is induced in a magnetic field applied along the $c$ axis of $H_{\mathrm{c,crit}} = 14$ T. A similar spin flop mechanism was detected for the magnetization along the $b$ axis at $H_{\mathrm{b,crit}} = 21$ T[13,14]. Then, the $b$ axis becomes the easy magnetization axis, similar to the magnetic behavior of FM URhGe. Here the so-called intermediate $b$ axis is characterized by a magnetic moment re-orientation at a critical magnetic field $H_R = 12$ T which restores the SC state[5, 15]. Recent papers revealed strong tricritical fluctuations in the vicinity of $H_R$[16, 17] accompanied by a Lifshitz-type transition and enhancement of the coefficient $\gamma$ [18-20].

We studied the magnetic properties and quantum critical phenomena in the UIr$_{1-x}$Rh$_x$Ge system which has an interesting FM/ antiferromagnetic (AFM) boundary at low temperature. This AFM/FM boundary is of interest because the AFM and FM are separated at this point throughout the whole orthorhombic TiNiSi-type U$T$Ge system by the Hill limit[4]. Many studies have been conducted to determine the delicate balance of magnetic interactions in U$T$Ge alloy systems, but they have been primarily on polycrystalline samples where the crucially important magnetocrystalline anisotropy remains hidden[4, 21-26]. Our investigation of single crystal study has allowed us to develop a general picture of the magnetism in the AFM part of the UIr$_{1-x}$Rh$_x$Ge system which surprisingly preserves many of the magnetic features of the FM parent compounds URhGe and UCoGe. Our discussion and conclusions are based on a detailed analysis of the crystal structure, magnetization, and heat capacity.

## II. Experimental

High-quality single crystals were grown by Czochralski pulling in a tetra arc furnace from polycrystalline precursors of nominal concentrations listed in Table I. Pulling speeds of 6 mm/h was used for the alloy compounds. The single crystals were several-centimeter long cylinders of 2-3 mm diameter. The pulled crystals were wrapped in Ta foil, sealed in quartz tubes under high vacuum, and annealed 14 days at 1000°C. The residual resistivity ratio (*RRR*) was substantially increased by this annealing procedure in the case of UIrGe from 2-4 up to several tens[27]. The *RRR* of the substituted compounds remains unchanged as observed in the other systems[4, 12, 21, 24, 25, 28, 29]. A precision spark erosion saw was used to cut appropriately shaped samples. An electron-probe microanalyser EPMA JXA-8900 (JEOL) has been used for the chemical analysis. Structural characterization was performed by single crystal x-ray diffraction using a Rigaku Rapid diffractometer. The recorded patterns were evaluated using ShelX software. The temperature and field dependent magnetization was measured along the principal crystallographic directions down to $T = 1.8$ K in applied magnetic fields up to 7 T using a commercial magnetometer (Magnetic Property Measurement System) MPMS 7T and 5T (Quantum Design). The heat capacity measurements were carried out down to 1.8 K with applied magnetic fields up to 9 T using a commercial Physical Property Measurement System PPMS (Quantum Design, DynaCool).

## III. Experimental results
### A. Chemical analysis

The microprobe analysis of all the alloying single crystals revealed a higher concentration of Ir than the nominal composition of the melt. Table I summarizes the nominal concentrations and the results of electron microprobe analyses. This disproportion causes a weak gradient of the Ir-Rh ratio along the single crystal ingots. The upper parts of the ingots are richer in Ir. The Rh concentration increases toward the bottom due to the prior consumption of Ir during the growth process. Detailed chemical analysis has revealed a weak gradient of the Ir-Rh ratio ~1 at.% (almost the detection limit of the method used) along the 40-mm-long ingot.

TABLE I. Chemical analyses of the studied single crystals in the $UIr_{1-x}Rh_xGe$ system. *Block 4 in Fig. 1. The compositions obtained by electron microprobe analyses are used in the later text.

| Nominal concentration | Microprobe analysis |
|---|---|
| AFM $UIr_{0.50}Rh_{0.50}Ge$ | $UIr_{0.58}Rh_{0.42}Ge_{1.00}$ |
| *AFM $UIr_{0.37}Rh_{0.63}Ge$ | $UIr_{0.45}Rh_{0.55}Ge_{0.98}$ |
| FM $UIr_{0.35}Rh_{0.65}Ge$ | $UIr_{0.43}Rh_{0.57}Ge_{0.99}$ |
| FM $UIr_{0.10}Rh_{0.90}Ge$ | $UIr_{0.14}Rh_{0.86}Ge_{0.99}$ |

We cut the single crystal of composition $UIr_{0.45}Rh_{0.55}Ge$, expected to have a robust AFM phase, into ~2-mm-long blocks and measured for each one the temperature dependent magnetization along the $c$ axis. Upper blocks 1–13 show robust AFM with Néel temperature determined from the magnetization maxima. Nonetheless, due to the weak Rh gradient we found the first signature of the nascent FM phase as a broad hump with roughly $T_C \approx 6.5$ K in block 14 of $UIr_{0.44}Rh_{0.56}Ge$. Simultaneously, the weak local maximum of the AFM phase still remains, fixed at $T_N \approx 3.9$ K (see Fig. 1).

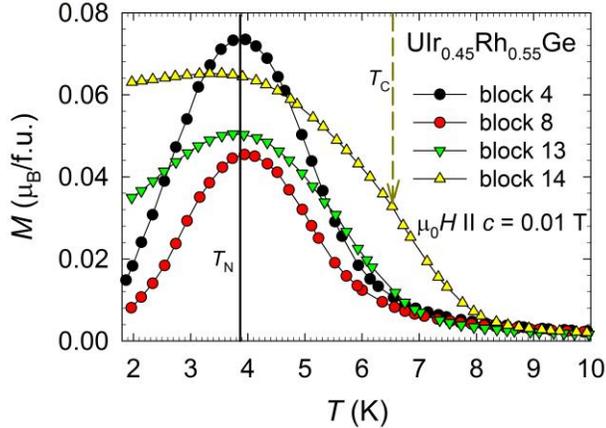

FIG. 1 (Color online) Temperature dependent magnetization of selected blocks of the $UIr_{0.45}Rh_{0.55}Ge$ single crystal. Increasing block numbers correspond to the direction from the neck to the end of the single crystal. The black line marks the lowest detectable $T_N$. The yellow dashed arrow marks the $T_C$ of the nascent FM phase. (f.u. = formula unit)

For the later research we used block 4 of composition $UIr_{0.45}Rh_{0.55}Ge$ as the ultimate AFM compound. The weak Rh-Ir gradient in the other single crystals does not have any noticeable effect, causing only a tiny shift of the ordering temperature of the robust FM or AFM phase. One unique block of length ~2 mm was always extracted from each single crystal and used for all experiments to avoid any effect of the gradient.

## B. XRD characterization

The high quality of each single crystal was verified by Laue patterns showing sharp reflections. Structural analysis by single crystal x-ray diffraction confirmed the orthorhombic TiNiSi-type structure and space group *Pnma* throughout the whole series. Results are summarized in Fig. 2 and the Appendix.

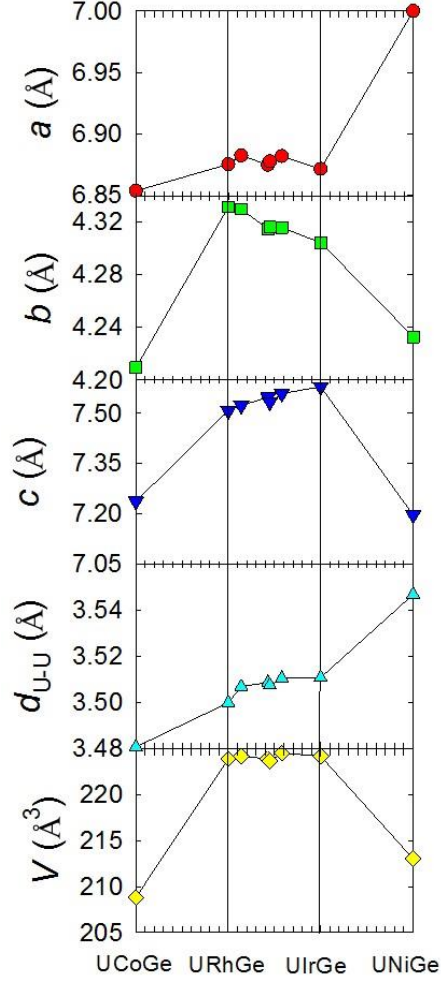

FIG. 2 (Color online) Lattice parameters in the $UIr_{1-x}Rh_xGe$ system as a function of concentration. Two neighboring compounds UCoGe and UNiGe are also plotted to show the gradually increasing nearest uranium-uranium distance $d_{U-U}$ utilized in the Hill plots[4]. Refined structural parameters for the $UIr_{1-x}Rh_xGe$ system are available in the Appendix.

The AFM/FM boundary in the $UIr_{1-x}Rh_xGe$ system is interesting from the standpoint that AFM UIrGe and FM URhGe have very similar lattice parameters arising from almost identical radii of the transition element ions[30]. The lattice similarity is evident when the $UIr_{1-x}Rh_xGe$ data are plotted together with the neighboring UCoGe and UNiGe showing the growing $d_{U-U}$ (Fig. 2). The unit cell volume of UIrGe is only about 0.07 % larger than that of URhGe[6, 31]. The very small change of the unit cell volume arises from nearly perfect cancelation of the weakly expanded lattice parameter *c* and shortened *b*. The crucial parameter *a* remains unchanged and reflects an almost constant $d_{U-U}$ distance. The shortening of the *b* axis reduces the second nearest $d_{U-U}$ distance from 3.758 Å (URhGe) to 3.747 Å (UIrGe), the zigzag chains separation being at distance *b*. The *bc* plane of the TiNiSi-type structure can be considered as derived from a deformed hexagonal lattice. Angle 70.39° characterizes the lattice of the URhGe in compared with the 70.09(5)° of UIrGe. The larger *c* of UIrGe sharpens the angle of the zigzag chains from 157.34(6)° of URhGe to 156.23(9)°. However, these variations do not lead to obvious conclusions concerning the FM/AFM boundary in the UIr$_{1-x}$Rh$_x$Ge based on a simple structural analysis. Moreover, the lattice parameters *b* and *c* develop unsystematically from UIrGe to another AFM UNiGe (Fig. 2).

### C. Magnetization

Magnetic ordering temperatures of all studied compounds were determined (Fig. 3, Table II). In the case of the parent UIrGe with the sharp AFM transition, we found clear agreement with $T_N$ estimated as a position of the sudden drop in the electrical resistivity[27], the peak maximum in the temperature dependent susceptibility along the *c* axis, and onset of the $\lambda$-anomaly in the heat capacity[14]. The identical procedure is applicable for the electrical resistivity and heat capacity of the FM URhGe[31]. Thus, we strictly followed this procedure for all alloy compounds studied. The estimation method used for $T_N$ as the maximum susceptibility also respects the method in the original polycrystalline paper[32]. We found two FM compounds, UIr$_{0.14}$Rh$_{0.86}$Ge and UIr$_{0.43}$Rh$_{0.57}$Ge, with Curie temperatures $T_C$ = 9.1 K and 6.2 K, respectively. The plateau of the weakly decreasing $T_C$ is broken by an acute fall to the Néel temperature $T_N$ = 3.9 K with the first AFM composition UIr$_{0.45}$Rh$_{0.55}$Ge. We particularly note that the FM/AFM boundary arises with an infinitesimal concentration step of the substituent elements. Magnetization studies of the gradient crystals have never found $T_C$ and $T_N$ to merge continuously. A further increase of Ir concentration is accompanied by growth of $T_N$. A $T_N$ = 7 K was found in UIr$_{0.58}$Rh$_{0.42}$Ge, increasing to the highest value $T_N$ = 16.5 K of parent UIrGe (Fig. 3). Ordering temperatures together with all the magnetic constants are summarized in Table II.

There is a discrepancy between our results and the original paper. UIr$_{0.45}$Rh$_{0.55}$Ge was reported as FM[32] while we still see clear AFM order. We will show below that a magnetic field of 0.1 T[32] along the *c* axis was strong enough to initiate the spin-flop transition.

Magnetic anisotropy of the AFM UIr$_{0.58}$Rh$_{0.42}$Ge still points to a simple collinear magnetic structure with magnetic moment aligned along the *c* axis. In contrast, a reproducible drop of magnetization is detected along the *a* and *b* axes at $T_N$ in UIrGe (Fig. 4). This may be evidence of canting of the magnetic moments as predicted by neutron diffraction close to the parent UIrGe.

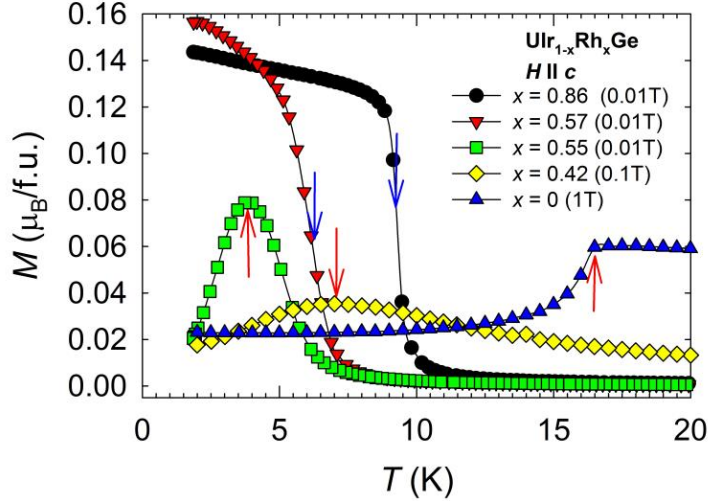

FIG. 3 (Color online) Temperature dependent magnetization of all studied compounds in the UIr$_{1-x}$Rh$_x$Ge system. The blue arrows indicate Curie temperatures, and the red arrows Néel temperatures. The curve for UIr$_{0.58}$Rh$_{0.42}$Ge was multiplied by 5 and the curve for UIrGe by 3 for clarity because of the reduced magnetic moments compared with the FM members. Curie temperatures were taken as the inflection points in the magnetization curves whereas and the Néel temperatures as the maxima of the peaks.

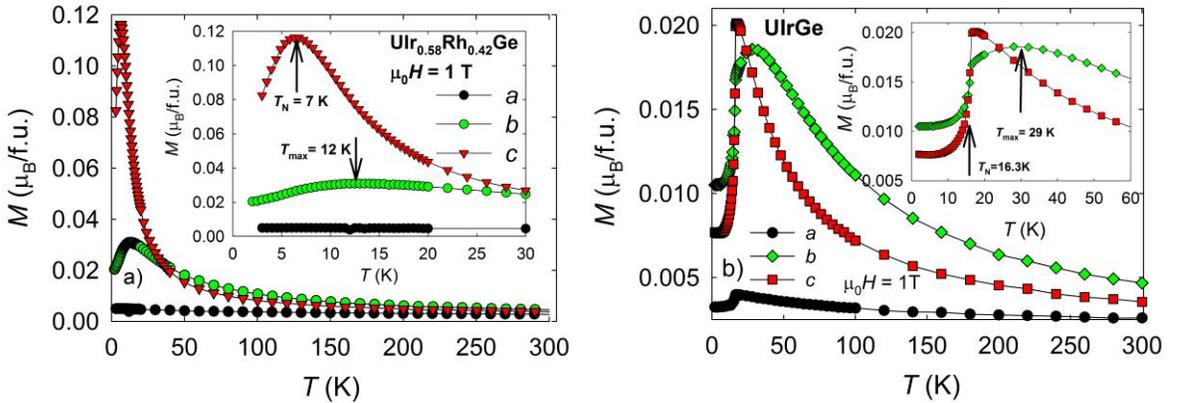

FIG. 4 (Color online) Temperature dependent magnetization of AFM UIr$_{0.58}$Rh$_{0.42}$Ge (a) and UIrGe (b) along all three crystallographic axes.

Current knowledge about the magnetic structure of UIrGe is quite unclear. Neutron diffraction on a single crystal[8, 33] suggested a component of magnetization along *a* which was not confirmed experimentally by magnetization up to 50 T[14]. Magnetic moment components in the *bc*-plane were found for the isostructural AFM UNiGe and UPdGe[9]. The UIrGe state resembles the *a* axis component of the magnetic moment proposed in UNiGe, although the *bc*-plane is magnetically soft with a complex magnetic phase diagram[34, 35].

Temperature dependent magnetization curves along the $b$ axis show complex behavior. A broad maximum is located above $T_N$ at $T_{max}$ = 11, 12, and 29 K in $UIr_{0.45}Rh_{0.55}Ge$, $UIr_{0.58}Rh_{0.42}Ge$ and $UIrGe$, respectively (Fig. 4). $T_{max}$ is also observable as a sharp peak in the FM $UIr_{0.14}Rh_{0.86}Ge$ (Fig. 5a) identical to parent $URhGe$[20] and as a broad peak in $UIr_{0.43}Rh_{0.57}Ge$ (Fig. 5b). A general feature is that $T_{max} \approx T_C$ for all FMs. In contrast, $T_{max} > T_N$ for all AFMs (see Table II). $T_{max}$ is shifted to lower temperatures with an increasing magnetic fields along the $b$ axis, demarking closed domes whose summits are located at higher field than those available to our magnetometer. Data from Figs. 5 and 6 will be used below for construction of the $H$-$T$ phase diagrams.

The temperature dependent inverse magnetic susceptibilities are strongly nonlinear up to 400 K. We had to use a modified Curie-Weiss law[36] which gives good agreement with the data in the interval ~30-400 K. Calculated magnetic constants for all materials are summarized in Table II.

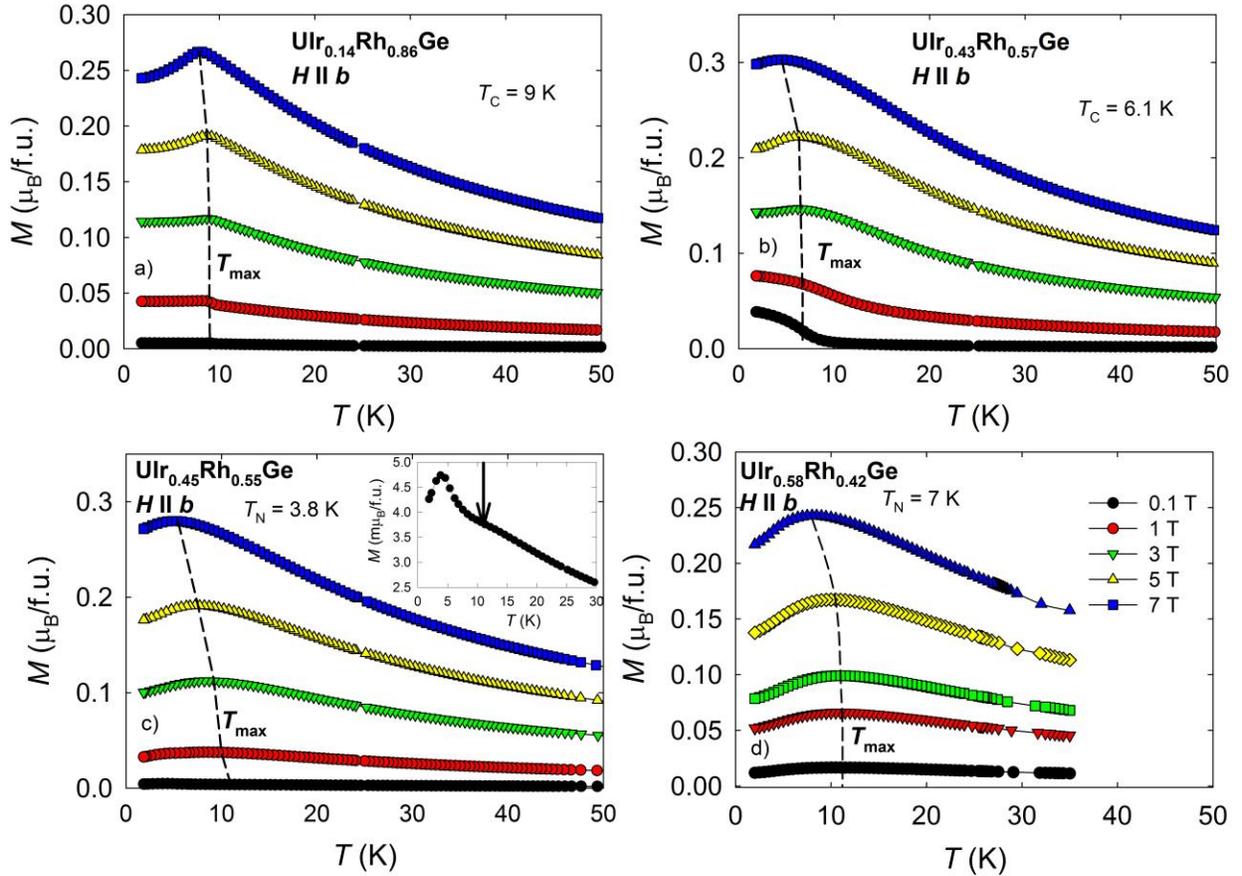

FIG. 5 (Color online) Temperature dependent magnetization of all studied compounds along the $b$ axis. The dashed lines tentatively mark the position of $T_{max}$ as a function of magnetic field. The inset in panel c shows the curve at 0.1 T in detail. $T_{max}$ is detected as a broad maximum. The peak at ~4 K is a projection of the easy axis due to a small misalignment of the sample.

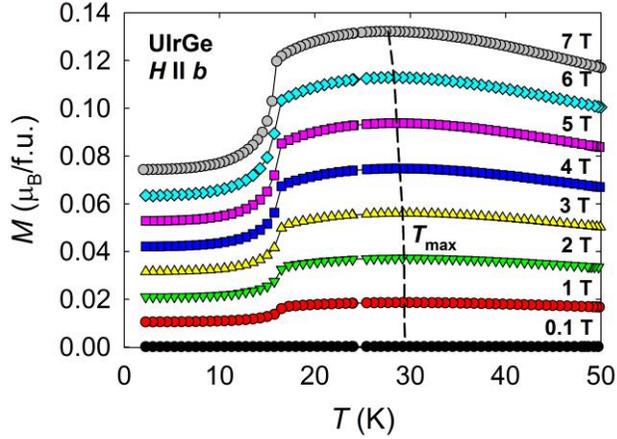

FIG. 6 (Color online) Temperature dependent magnetization of UIrGe in a series of magnetic fields applied along the $b$ axis. The dashed line tentatively marks the position of $T_{max}$ as a function of the magnetic field.

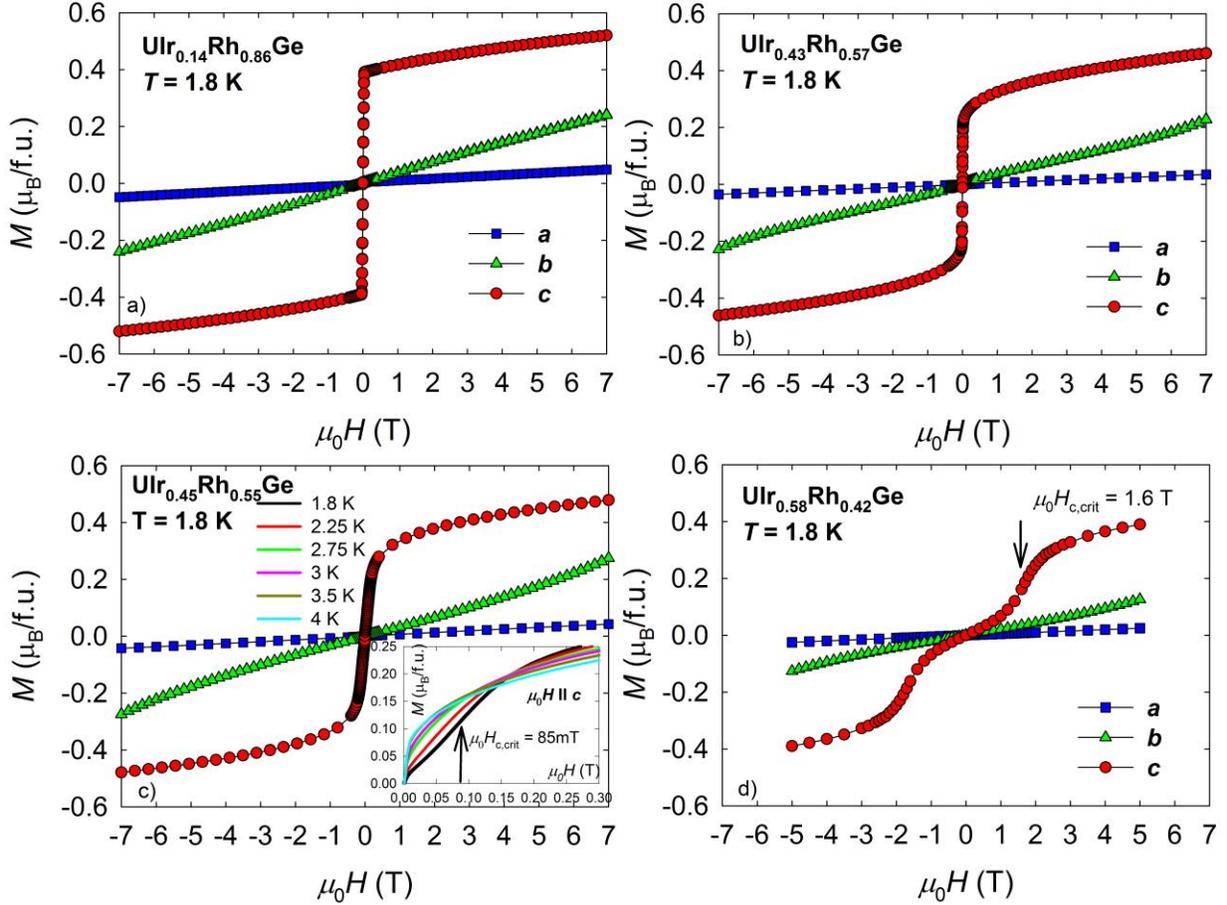

FIG. 7 (Color online) Magnetization isotherms of the alloy compounds along all three crystallographic axes. The value of $H_{c,crit}$ is taken as the inflection point of the metamagnetic jumps.

Effective magnetic moments are reduced in compared with free $U^{3+}$ and $U^{4+}$ ionic values in all compounds along all three axes. Magnetization isotherms show the hard magnetization *a* axis. The easy magnetization axis is the *c* axis. The spontaneous magnetic moment of the FM compounds $\mu_{sp}$ gradually decreases with increasing Ir content having almost half the value at the AFM/FM boundary of that of parent URhGe (Table II). Hysteresis of the FM compounds at temperature 1.8 K is significantly suppressed to a value of only ~0.001 T.

A metamagnetic jump instantly appears in the first AFM compound $UIr_{0.45}Rh_{0.55}Ge$ at the critical field of the spin-flop transition $\mu_0 H_{c,crit} = 0.085$ T (Fig. 7c - inset). The metamagnetic transition clearly disappears at $T_N$ which is strong evidence of the intrinsic bulk AFM. The $H_{c,crit}$ of $UIr_{0.45}Rh_{0.55}Ge$ is lower than that found in previous paper, which considered this compound to be FM. Approaching UIrGe strengthens $H_{c,crit}$ up to a final value of 14 T (Table II).

Magnetic moment re-orientation along the *b* axis is a strongly studied phenomenon of URhGe because of the magnetic field-induced SC[37]. A similar spin-flop transition also appears at $\mu_0 H_{b,crit} = 21$ T in UIrGe. We found the value of the critical field $H_{b,crit}$ to be above the limit of a common superconducting interference device (SQUID) magnetometer in $UIr_{1-x}Rh_xGe$. The metamagnetic transition can be inferred only just at the AFM/FM boundary by the tenuous increase of magnetization at the maximum available field (Figs. 7b and 7c). A large value of $H_{b,crit}$ evidently passes through the AFM/FM boundary, which will be established later by the heat capacity method.

In contrast to the *c* axis, the value of the magnetization along the *b* axis grows even above $T_N$ in the AFM compounds. Magnetization isotherms are characterized by convex curvature indicative of an additional metamagnetic transition existing above $T_N$. The maximum magnetization and linear character of the magnetization isotherms are reached at temperature $T_{max}$ (see the example in Fig. 8), which raises the question of whether the metamagnetic jump along the *b* axis is associated with $T_N$ or $T_{max}$. We will solve this issue later in our discussion of the *H-T* phase diagrams.

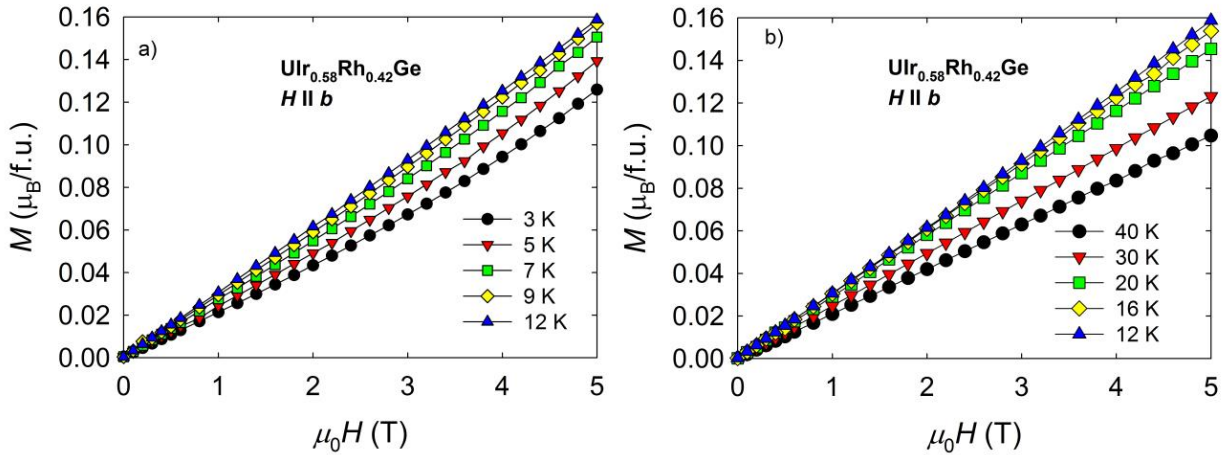

FIG. 8 (Color online) Representative magnetization isotherms along the *b*-axis. Maximum magnetization of $UIr_{0.58}Rh_{0.42}Ge$ ($T_N = 7$ K) along the *b* axis was reached at 12 K corresponding to $T_{max}$. The isotherms have Brillouin character above this characteristic temperature.

TABLE II. Magnetic and thermodynamic constants of all studied compounds in the UIr$_{1-x}$Rh$_x$Ge system. A dashed line separates the FM and AFM compounds. Nonetheless, evidence of the nascent FM phase was detected by heat capacity data in $x = 0.55$. Magnetic constants of URhGe with rather limited agreement are available in Ref.[31] by fitting the Curie-Weiss law compared with our fitting by modified Curie-Weiss law. *Values were estimated with assumption of a constant $H_{b,crit}/T_C$ ratio deduced from the observed experimental values for FM URhGe and UIr$_{0.43}$Rh$_{0.57}$Ge. Dashed cells-physical quantity is not defined for given material. Empty cells-the value of the physical quantity was not established or established by different model than we used in our work.

| UIr$_{1-x}$Rh$_x$Ge | x = 1 | x = 0.86 | x = 0.57 | x = 0.55 | x = 0.42 | x = 0 |
|---|---|---|---|---|---|---|
| $T_C$ (K) | 9.5 | 9.1 | 6.2 | - | - | - |
| $T_N$ (K) | - | - | - | 3.9 | 7 | 16.5 |
| $T_{max}$ (K) | = $T_C$ | = $T_C$ | = $T_C$ | ~11 | 12 | 29 |
| $T_{max}/T_C$ ($T_N$) | 1 | 1 | 1 | 2.8 | 1.71 | 1.75 |
| $\mu_0 H_{c,crit}$ (T) | - | - | - | 0.085 | 1.6 | 14[14] |
| $\mu_0 H_{b,crit}$ (T) | 12.5 | 12.1* | 8.3 | 6.6 | 7.7 | 21[14] |
| $\mu_0 H_{b,crit}/T_C$ ($T_N$) | 1.32 | 1.33* | 1.33 | 1.69 | 1.1 | 1.27 |
| $\mu_0 H_{b,crit}/T_{max}$ | 1.32 | 1.33* | 1.33 | 0.60 | 0.64 | 0.72 |
| $\mu_{eff}$ (a) ($\mu_B$/f.u.) | | 1.55 | 1.51 | 1.42 | 1.29 | 1.02 |
| $\mu_{eff}$ (b) ($\mu_B$/f.u.) | | 2.12 | 2.12 | 2.11 | 2.3 | 2.52 |
| $\mu_{eff}$ (c) ($\mu_B$/f.u.) | | 1.78 | 1.73 | 1.73 | 1.67 | 1.66 |
| $\mu_{sp}$ ($\mu_B$/f.u.) | 0.43[31] | 0.39 | 0.24 | - | - | - |
| $\theta_p$ (a) (K) | | -109 | -131 | -118 | -112 | -97 |
| $\theta_p$ (b) (K) | | -17.2 | -11.1 | -13.7 | -20.3 | -34.1 |
| $\theta_p$ (c) (K) | | 5.5 | 3.4 | 4.5 | 3.9 | -10 |
| $\chi_0$ (a) ($10^{-8}$mol/m$^3$) | | 1.19 | 1.17 | 1.18 | 1.33 | 1.39 |
| $\chi_0$ (b) ($10^{-8}$mol/m$^3$) | | 0.88 | 0.91 | 0.88 | 0.58 | 0.25 |
| $\chi_0$ (c) ($10^{-8}$mol/m$^3$) | | 1.23 | 1.19 | 1.18 | 1.36 | 1.10 |
| $\gamma$ (mJ/mol K$^2$) | 163[20] | 160 | 175 | ~120 | ~70 | 16 |
| $\Delta C_p/T$ (mJ/mol K$^2$) | 200[12] | 180 | 75 | ~20 | ~80 | 750 |
| $S_{mag}$ ($R \ln 2$) | 0.2[31] | 0.17 | 0.073 | | | 0.23 |

### D. Heat capacity

Heat capacities of the FM compounds are characterized by a clear $\lambda$-type anomaly (Figs. 9a and b). The shape of the anomalies abruptly changes to a broad maximum on the AFM side, which gradually transforms into the $\lambda$-type anomaly of pure UIrGe (Figs. 9c-f and Fig. 10). Nevertheless, the rapid drop of the heat capacity below $T_N$ typical for all AFM UIr$_{1-x}$Rh$_x$Ge is maintained, even though the UIr$_{0.45}$Rh$_{0.55}$Ge peak close to the boundary is rather weak. The ordering temperatures were established as the onsets of the peaks and are consistent with the results of magnetization data. UIr$_{0.45}$Rh$_{0.55}$Ge also provides evidence of the discontinuity between the $T_C$ and $T_N$ (Figs. 9c and d). We detected here a portion (~15 %) of the nascent FM phase with ordering temperature ~6 K while the predominant AFM phase has clearly lower $T_N$ = 3.7 K with no sign of merging.

Magnetic anomalies along the *c*- axis rapidly vanish in a magnetic field in all compounds. Significantly larger magnetic field must be applied along the *b* axis (Fig. 9).

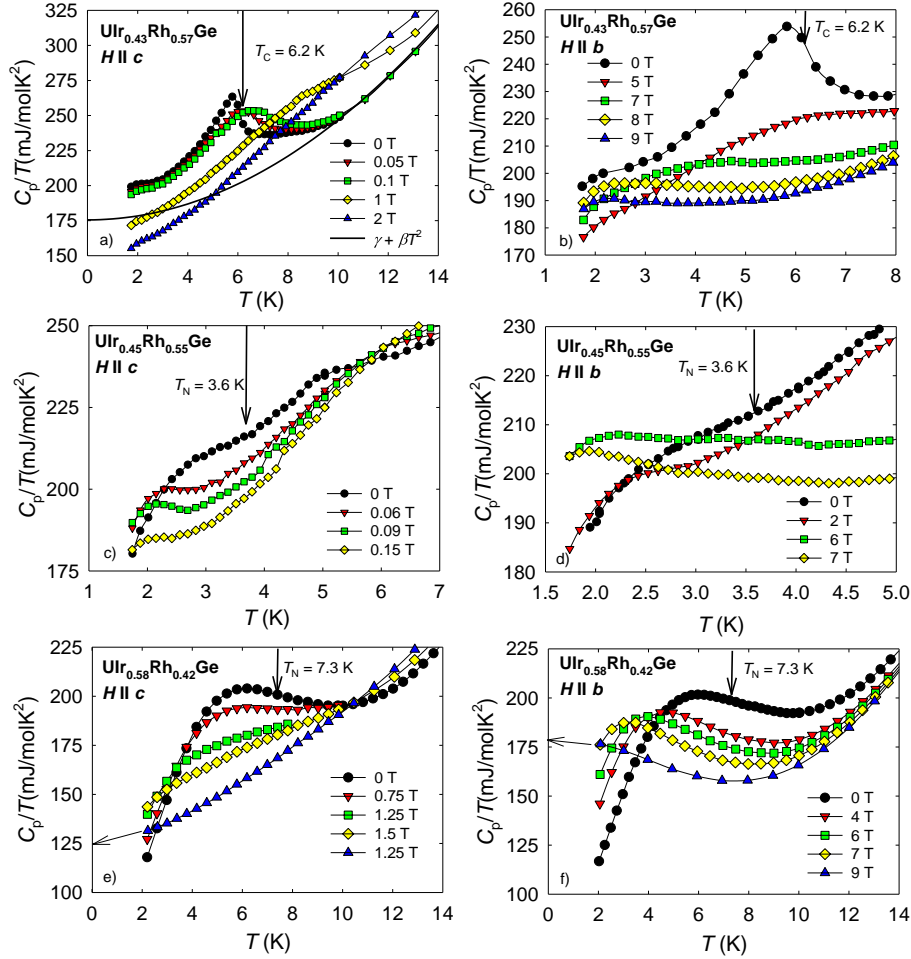

FIG. 9 (Color online) Heat capacity of UIr$_{0.43}$Rh$_{0.57}$Ge, UIr$_{0.45}$Rh$_{0.55}$Ge and UIr$_{0.58}$Rh$_{0.42}$Ge. Left panels show data in the magnetic field applied along the *c* axis, and the right panels show data in the magnetic field applied along the *b* axis. Black arrows mark the positions of the ordering temperatures. Two broad maxima are detected in the UIr$_{0.45}$Rh$_{0.55}$Ge data. The upper maximum is a result of the nascent FM phase, and the bottom maximum at 3.6 K represents the dominant AFM phase. We estimated the volume of the FM phase ~15% by considering the considerably reduced FM peak in compared with pure FM UIr$_{0.43}$Rh$_{0.57}$Ge. There is also evidence of the restoration of the peak in the FM UIr$_{0.43}$Rh$_{0.57}$Ge in panel b at 2.5 K and 9 T. The originally very broad maximum transforms to a narrow peak in UIr$_{0.43}$Rh$_{0.57}$Ge at ~2 K and in the magnetic field 0.09 and 7 T along the *c* and *b* axes visible in panels c and d. The effect is less clear in UIr$_{0.58}$Rh$_{0.42}$Ge.

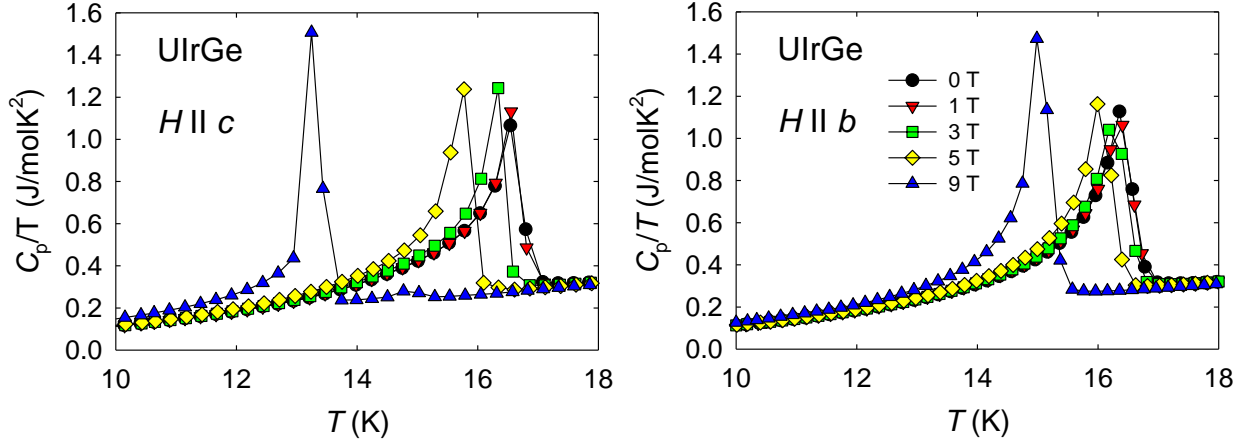

FIG. 10 (Color online) Heat capacity of UIrGe. (a) Data with magnetic field applied along the $c$-axis, and (b) data with magnetic field applied along the $b$-axes. There is evidence of narrowing and increase of peak height $\Delta C_p/T_{\text{UIrGe}}$ at the maximum magnetic field in both panels.

We extracted the phonon parts $C_{\text{ph}}$ by an identical procedure used previously for URhGe and UIrGe[6, 31]. The Debye model and the low temperature expression $C_p/T = \gamma + \beta T^2$ are used to analyze the experimental data, giving similar Debye temperatures. The calculated magnetic entropy $S_{\text{mag}}$ is reduced from $0.2R\ln 2$ in URhGe downwards to $0.073R\ln 2$ in the ultimate FM UIr$_{0.43}$Rh$_{0.57}$Ge (Table II). Calculated Debye temperatures of 198 K and 202 K using the formula $\theta_D^3 = 3(12\pi^4 R/5\beta)$ (where $R$ is the gas constant and $\beta$ phonon coupling constant), of FM UIr$_{0.14}$Rh$_{0.86}$Ge and UIr$_{0.43}$Rh$_{0.57}$Ge are similar to those of the parent compounds URhGe and UIrGe.

We point out a discrepancy with the previously published heat capacity data and the $\Delta C_p/T_{\text{UIrGe}}$ parameter. A significantly lower and broader peak is reported by Prokes et al.[6] and Ramirez et al.[7] and a double peak anomaly probably due to a parasitic grain in Chang et al.[38] in comparison to S. Yoshii et al.[14]. Our experimental observation $\Delta C_p/T_{\text{UIrGe}} \approx 750$ mJ/molK$^2$ on a high quality single crystal (residual resistivity ratio-$RRR$ 36) is in agreement with S. Yoshii et al. However, $S_{\text{mag}} = 0.23R\ln 2$ is still in agreement with the Prokes et al. paper[6] due to the much narrower character of the peak. It seems that the high-quality samples narrow and increase the heat capacity peak, but magnetic entropy $S_{\text{mag}}$ remains conserved. $S_{\text{mag}}$ of UIrGe is the same as for URhGe although $\Delta C_p/T$ is almost four times larger. The high $\Delta C_p/T_{\text{UIrGe}}$ is presumably the result of the opening of a large AFM gap at $T_N$[6].

A specific feature of the heat capacity is development of peak shapes of the transitions close to magnetic field $H_{\text{crit}}$, particularly in UIrGe. The original $\lambda$-type anomaly associated with the second order transition is transformed to a sharp peak of a first order-like transition of the considerably larger jump in $\Delta C_p/T_b^{14\text{T}}{}_{\text{UIrGe}} \approx 2600$ mJ/mol K$^2$. Our experimental observations (Fig. 10) are in agreement with the data in Ref.[39]. The clear $\lambda$-type peak in the heat capacity data of FM UIr$_{0.43}$Rh$_{0.57}$G begins broad in the magnetic field along $b$. However, the peak restores at ~2.5 K and magnetic field 8 and 9 T (Fig. 9b). Narrow peaks are also developed in the AFM UIr$_{0.45}$Rh$_{0.55}$Ge (Fig. 9c and d) at 2 K and magnetic field close to $H_{\text{crit}}$ along the $b$- and $c$- axes. The effect was not was observable in UIr$_{0.58}$Rh$_{0.42}$Ge (Fig. 9e and f).

We use heat capacity to estimate the value of the $H_{crit}$, which should be clearly detectable, in the thermodynamic Maxwell relation

$$\left(\frac{\partial S}{\partial H}\right)_T = \left(\frac{\partial M}{\partial T}\right)_H \quad (1)$$

Assuming the Fermi liquid state with $\sim T^2$ dependence of M, one can obtain the field derivative of $\gamma$ by the differentiation of Eq.1 with respect to temperature.

$$\left(\frac{\partial \gamma}{\partial H}\right)_T = \left(\frac{\partial^2 M}{\partial T^2}\right)_H = 2\beta \quad (2)$$

We experimentally performed field dependent scans of the heat capacity within this theoretical approach (see the results in Fig. 11). Because of finite temperature we plot the results as $C_p/T$. The using of a commercial PPMS $^3$He heat capacity puck down to 0.4 K was impossible because of the strong mechanical force of the highly anisotropic samples in the magnetic field applied along the hard magnetization axes. $H_{crit}$ is another quantity supporting the discontinuous AFM/FM boundary. A step of ~1.7 T was observed between the $H_{b,crit}$ values of FM UIr$_{0.43}$Rh$_{0.57}$Ge and AFM UIr$_{0.45}$Rh$_{0.55}$Ge. The magnetic field dependent heat capacity isotherms are also a tool to uncover the intrinsic character of the magnetic ground state of the each compound, particularly in the vicinity of the AFM/FM boundary. A decreasing character of $C_p/T$ is observed only in the FM compounds in magnetic field applied along the *c*-axis (Fig. 11a). In contrast, maxima corresponding to $H_{c,crit}$ are seen in the heat capacity isotherms (Figs. 11c and e) of the AFM compounds in agreement with the magnetization data (Table II). The height of the maxima along the *c*-axis are approximately two/fifths of the height along the *b*-axis in both cases.

A curvature of the heat capacity isotherms along the *b*-axis is maintained even above $T_N$ (Figs. 11d and f). This trend substantially weakens at temperatures close to $T_{max}$. The recorded isotherms will be used for later construction of the *H-T* phase diagrams.

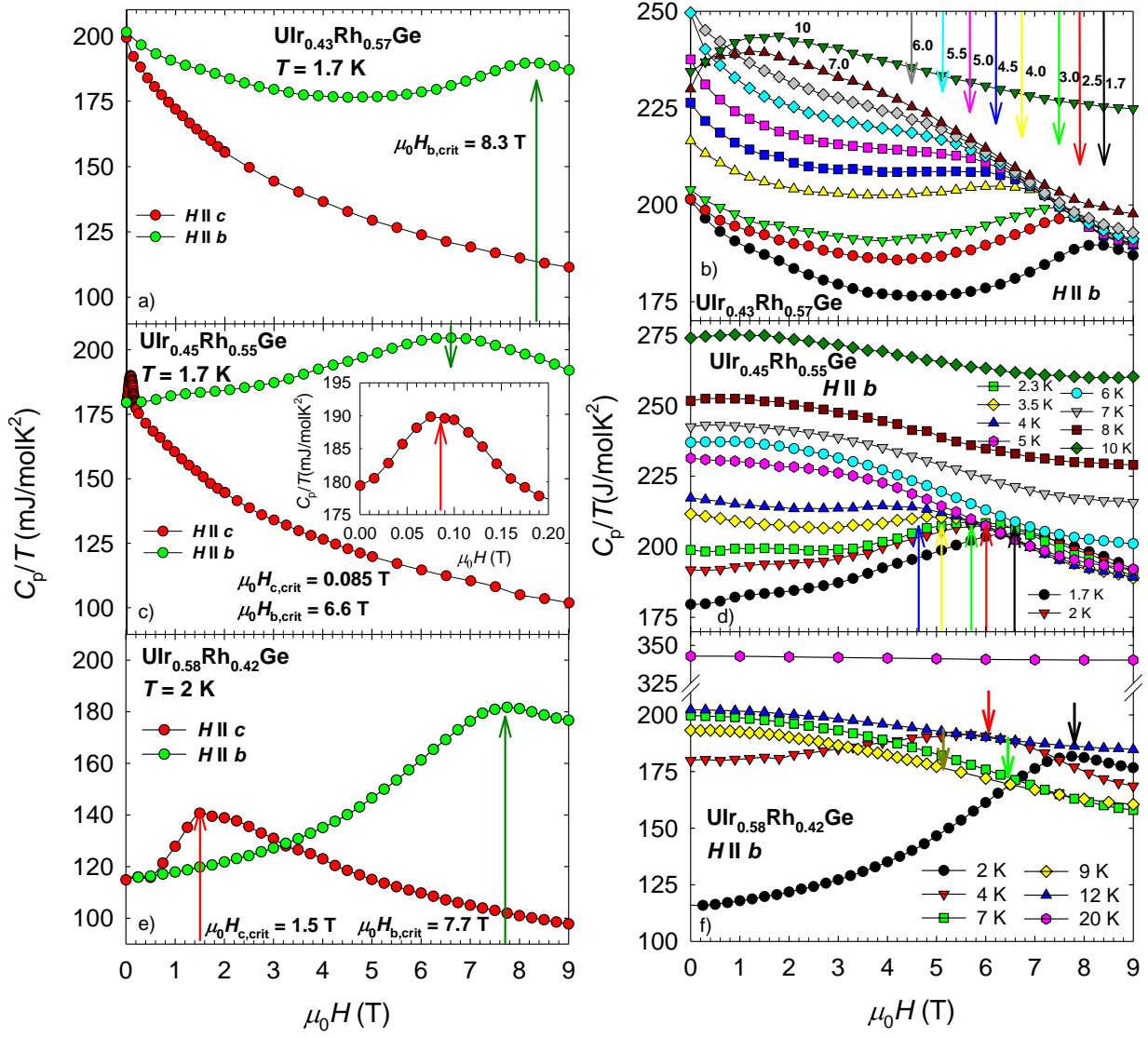

FIG. 11 (Color online) Field dependent experimental heat capacity data $C_p/T$ of UIr$_{0.43}$Rh$_{0.57}$Ge (a,b), UIr$_{0.45}$Rh$_{0.55}$Ge (c,d), and UIr$_{0.58}$Rh$_{0.42}$Ge (e,f) in external magnetic field applied along the $c$ and $b$ axes at various temperatures. The arrows point to the values of $H_{b,\text{crit}}$ used for later construction of the $H$-$T$ phase diagrams.

The FM UIr$_{1-x}$Rh$_x$Ge is characterized by an almost constant value of $\gamma$ with weak growth toward the boundary (Fig. 12). Here, the $\gamma$ coefficient suddenly falls and approaches $\gamma_{\text{UIrGe}} = 16$ mJ/mol K$^2$ (Table II).

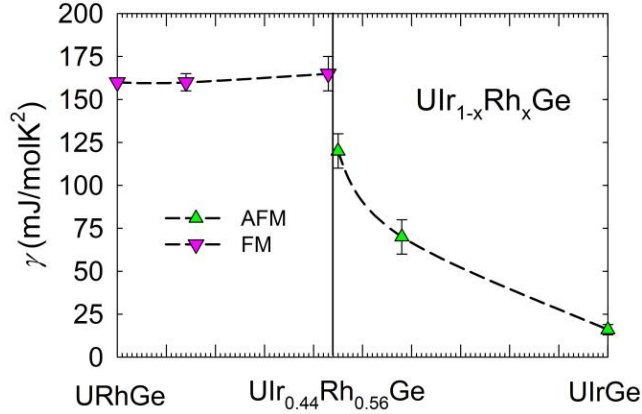

FIG. 12 (Color online) Evolution of the Sommerfeld coefficient $\gamma$ in the UIr$_{1-x}$Rh$_x$Ge system obtained by extrapolation of the data using $C_p/T = \gamma + \beta T^2$. $\gamma$ of AFMs UIr$_{0.45}$Rh$_{0.55}$Ge and UIr$_{0.58}$Rh$_{0.42}$Ge were estimated by tentative extension of the broad peaks to zero temperature resulting in larger error bars (Fig. 13).

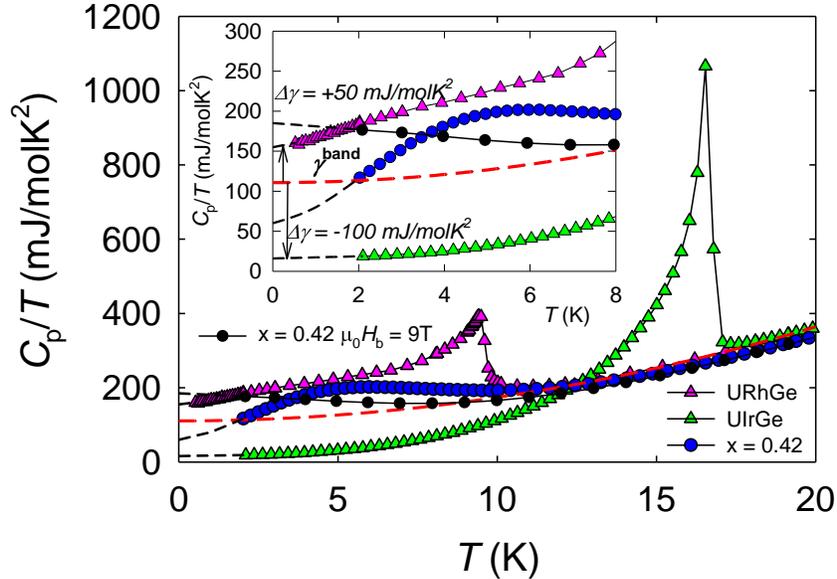

FIG. 13 (Color online) Temperature dependent $C_p/T$ of URhGe, UIrGe and UIr$_{0.58}$Rh$_{0.42}$Ge. The red line represents extrapolation of the paramagnetic part of the heat capacity of URhGe using $C_p/T = \gamma + \beta T^2$ to zero temperature, giving a value of $\gamma^{band}_{URhGe} \approx 110$ mJ/mol K$^2$. Other AFM compounds UIrGe and UIr$_{0.58}$Rh$_{0.42}$Ge are characterized by almost identical values of $\gamma^{band}$. The inset shows the low temperature interval. The value of $\gamma$ in the vicinity of $H_{b,crit}$ of the UIr$_{0.58}$Rh$_{0.42}$Ge compound $\gamma_b^{9T}{}_{UIr0.58Rh0.42Ge} \approx 180$ mJ/mol K$^2$ is apparently enhanced in compared with the value at zero field $\gamma_{UIr0.58Rh0.42Ge} \approx 70$ mJ/mol K$^2$ and $\gamma^{band}$.

The $\gamma$ coefficient is rather reduced in value compared with the FM SCs UCoGe and URhGe of $\gamma \approx 60$ and 160 mJ/mol K$^2$, respectively[3, 31]. On the other hand, extrapolation of the paramagnetic region of the $C_p/T$ data to zero temperature using $C_p/T = \gamma + \beta T^2$ points to a significantly larger value of $\gamma^{band}_{URhGe} \approx 110$ mJ/mol K$^2$ (Fig. 13). The $\gamma_{URhGe} \approx 160$ mJ/mol K$^2$ is about ~50 mJ/molK$^2$ higher than that of $\gamma^{band} \approx 110$ mJ/mol K$^2$ [20]. On the other hand $\gamma_{UIrGe} \approx 16$ mJ/molK$^2$ is about ~100 mJ/mol K$^2$ lower, indicative of a low density of 5$f$ states at $E_F$ probably due to opening of an AFM gap. We will discuss below renewal of magnetic fluctuations in the AFM compounds by magnetic field along the $b$- and $c$- axes.

## IV. Discussion
### A. The AFM/FM boundary in UIr$_{1-x}$Rh$_x$Ge

We collected magnetic parameters of all the studied compounds in the UIr$_{1-x}$Rh$_x$Ge system and constructed a magnetic phase diagram (Fig. 14). The remarkable result is confirmation of the discontinuity in all magnetic quantities between the utmost AFM UIr$_{0.45}$Rh$_{0.55}$Ge and FM UIr$_{0.43}$Rh$_{0.57}$Ge. We detected the discontinuity in the ordering temperatures as well as in $H_{b,crit}$ and a finite value of $H_{c,crit}$ on the AFM side. The discontinuity of the first order transition between the FM and AFM at critical concentration of ~UIr$_{0.44}$Rh$_{0.56}$Ge is also supported by the heat capacity of the AFM UIr$_{0.45}$Rh$_{0.55}$Ge with a nascent FM phase with no sign of a merger of $T_C$ and $T_N$. Instead, a clear gap of ~2.3 K was detected. A particularly important issue is the evolution of $T_{max}$. It is an intriguing property of URhGe where $T_C \approx T_{max}$ [15, 20] and this trend is maintained towards the ultimate FM compound UIr$_{0.43}$Rh$_{0.57}$Ge. $T_{max}$ suddenly splits from $T_N$ at the AFM border.

Consider the first order transition at the FM/AFM boundary; both $T_N$ and $T_C$ are finite at the boundary. Such a phase diagram could be realized in a system involving two independent magnetic intra- ($J$) and inter- ($J^*$) chain couplings along the $a$ axis. Indeed, U moments are aligned ferromagnetically along the chain in both UIrGe and URhGe. This indicates $J > 0$ (FM) on both sides of the boundary. The AFM/FM boundary can then be defined as the point where only $J^*$ changes sign from $J^* < 0$ (AFM for Ir) to $J^* > 0$ (FM for Rh). Because of the discontinuous transition $J^* \neq 0$. Naturally, both $T_N$ and $T_C$ are finite with $J > 0$.

Room temperature crystal structure analysis does not provide any clear solution for variation of the $J^*$-$J$ balance, but an abrupt change of the lattice parameters cannot be excluded especially since the thermal expansion coefficients $\alpha_i$ are not well-known for UIrGe around $T_N$ [40]. It is worth nothing here, that the UIrGe hydride is a ferromagnet of $T_C = 28$ K that coincides with the position of $T_{max}$ in the parent UIrGe [41]. However, the magnetocrystalline anisotropy, detailed crystal and magnetic structures of the UIrGe hydride are unknown to bring a light to the possible development of the $J$-$J^*$ balance.

The second scenario considers the effect of band width $W_d$ of the valence 4$d$ and 5$d$ states of Rh and Ir [42], respectively, affecting the 5$f$-$d$ hybridization and also the spin-orbit $s$-$o$ interaction of the much heavier Ir ion [43]. This should be verified by detail electronic structure study by ARPES or dHvA effect in UIrGe.

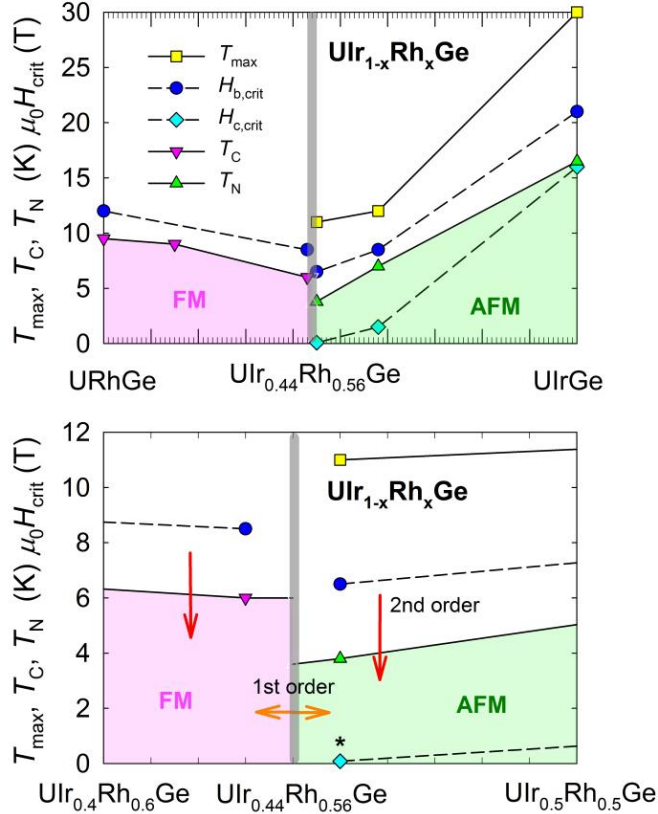

FIG. 14. (Color online) Magnetic phase diagram of the UIr$_{1-x}$Rh$_x$Ge system. (a) Full concertation range, and (b) area around the AFM/FM boundary in detail. The asterisk is 0.085T (Table II). The critical concentration is tentatively established at $x_{crit} = 0.56$. The lines are guides to the eye.

### B. QCP in the UIr$_{1-x}$Rh$_x$Ge system at $x_{crit}$

Evolution of the ordering temperature through the UCoGe-URhGe-UIrGe system is displayed in the $d_{U-U}$-$T$ phase diagram (Fig. 15). We used the parameter $d_{U-U}$ instead of common concentration $x$ because the phase diagram connects together two different alloy systems. Nevertheless, $d_{U-U}$ also is not a physically relevant parameter, and finding a better one is subject of further research.

Existence of quantum critical points (QCPs) were reported in the neighboring alloying FM-PM and AFM-paramagnetic (PM) systems URh$_{1-x}$Ru$_x$Ge[29, 44], UCo$_{1-x}$Fe$_x$Ge[28], UCo$_{1-x}$Ru$_x$Ge[4] or UPd$_{1-x}$Ru$_x$Ge[21]. The UIr$_{1-x}$Rh$_x$Ge system behaves like the other AFM/FM alloy system UPd$_{1-x}$Co$_x$Ge where magnetic order survives in the entire concentration range[22]. In contrast to UPd$_{1-x}$Co$_x$Ge, a deep local minimum in the ordering temperatures is created at the AFM/FM boundary in UIr$_{1-x}$Rh$_x$Ge almost at the level of $T_C$ of UCoGe. Secondly, the analysis suggests here an enhancement of the coefficient $\gamma_{UIr0.43Rh0.57Ge} \approx 175$ mJ/mol K$^2$ (Fig. 12), the highest in the UCoGe-URhGe-UIrGe system, indicating enhancement of the magnetic fluctuations typical for the development of a QCP reported in the above listed alloy systems. However, magnetic fluctuations are interrupted in UIr$_{1-x}$Rh$_x$Ge by a very stable AFM phase and a QCP is not realized.

In particular the evolution of the $\gamma$ coefficient in Fig. 12 confirms a sudden reconstruction of the electronic structure, probably by the AFM gap opening in the magnetic Brillouin zone.

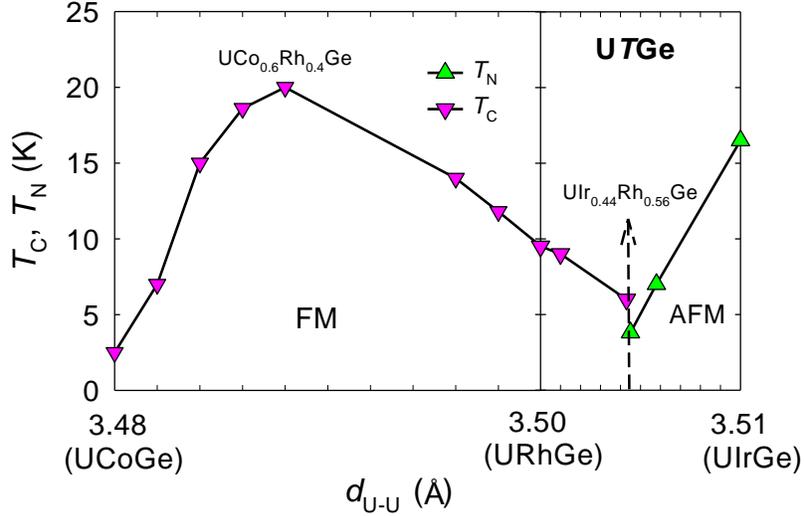

FIG. 15. (Color online) $d_{U-U}$-$T$ phase diagram of the UCoGe-URhGe-UIrGe system. The $UCo_{1-x}Rh_xGe$ panel is constructed based on data in ref.[23]. The width of the $UCo_{1-x}Rh_xGe$ and $UIr_{1-x}Rh_xGe$ panels corresponds to the nearest uranium ion distance $d_{U-U}$, assuming Vegard's law[45].

### C. *H - T* phase diagrams

$H_b$-$T$ phase diagrams of the FM compounds are displayed in Fig. 16. A gradual increase of Ir concentration in URhGe suppresses both $T_C$ and $H_{b,crit}$. Moreover, the phase diagram in Fig. 14 has suggested uniformly decreasing $T_C$ in the wide interval from ~$UCo_{0.6}Rh_{0.4}Ge$ down to $UIr_{0.43}Rh_{0.57}Ge$. Thus, the temperature dependence of an order parameter and energy scale of the magnetic interactions in all these compounds seem to be of the same nature as seen in the normalized phase diagram with overlap of all curves. It has also allowed us to tentatively draw the phase diagram of $UIr_{0.14}Rh_{0.86}Ge$ whose critical field $H_{b,crit}$ was higher than that of available magnetic fields in the instruments used.

The normalized phase diagram together with the recovered heat capacity anomaly of $UIr_{0.43}Rh_{0.57}Ge$ at low temperature and magnetic field close to $H_{b,crit}$ (Fig. 9b) raise a fundamental question concerning the development of the first order transition in the proximity of $H_{b,crit}$ through the ~$UCo_{0.6}Rh_{0.4}Ge$ - $UIr_{0.43}Rh_{0.57}Ge$ region characterized by a monotonous decrease of $T_C$. Recent studies of URhGe confirmed the transformation of the second order to a first order FM/PM transition at tricritical point (TCP) located at finite temperature with characteristic bifurcation to the wing-structure phase diagram[46]. Surprisingly, a similar wing-structure phase diagram was confirmed by a detailed NMR investigation of the alloying compound $UCo_{0.1}Rh_{0.9}Ge$[17], which is incorporated in Fig. 16. It is in contrast with the prediction for the FM/PM quantum phase transition quantum phase transition in the disordered FM metallic systems where a continuous second order phase transition is maintained down to zero temperature[47, 48].

$UCo_{1-x}Rh_xGe$ and $UIr_{1-x}Rh_xGe$ may represent a particular case where the first order transition is attainable in a magnetic field along the *b* axis at finite disorder strength because there are essential differences compared with the alloy systems with QCP. First, $UCo_{1-x}Rh_xGe$ and $UIr_{1-x}Rh_xGe$ represent alloying between isoelectronic transition metals. Another specific feature of the $UIr_{1-x}Rh_xGe$ system is the almost negligible variation of the ionic diameter of Rh and Ir reflected in a very weak change in the lattice parameters which inhibits a local structural disorder. Such closeness of the magnetic features was already observed in other Rh-Ir alloy systems[49, 50] but we simultaneously avoid generalization of the suggested scenario of the possible first order transition for the all types of isoelectronic alloy systems. Taking into account the scaling parameter $H_{b,crit}/T_C$, phase diagrams in Fig. 15 and the recovered heat capacity anomaly (Fig. 9b) in the vicinity of $H_{b,crit}$, we can assume the scenario of the first order transition for all the compositions in the region of ~$UCo_{0.6}Rh_{0.4}Ge$ - $UIr_{0.43}Rh_{0.57}Ge$.

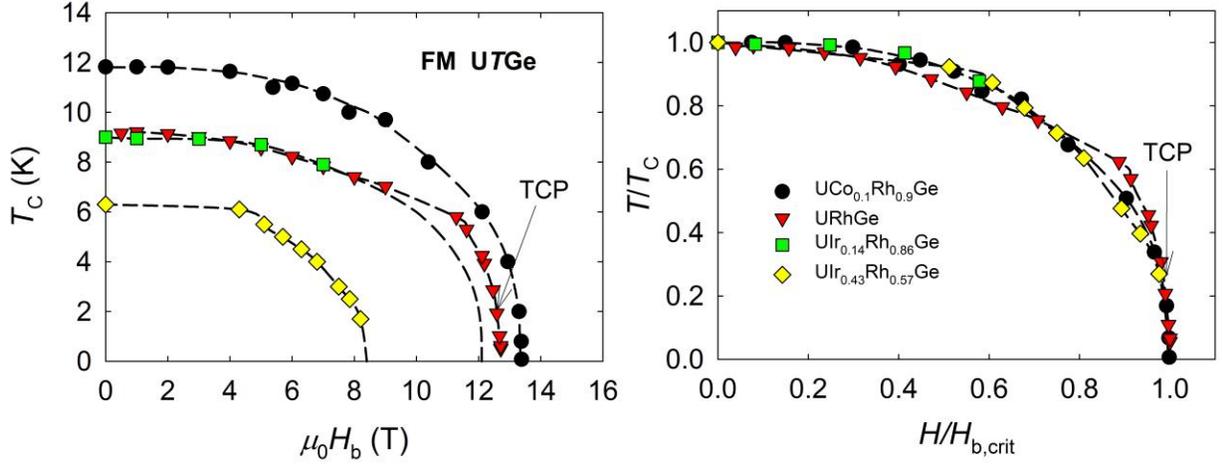

FIG. 16. (Color online) (a) $H_b$-$T$ phase diagram of the FM compounds in the $UIr_{1-x}Rh_xGe$ system extended about FM $UCo_{0.1}Rh_{0.9}Ge$[17] and URhGe[51]. Scaling of the magnetization data was used for construction of the phase diagram of $UIr_{0.14}Rh_{0.86}Ge$. (b) The normalized $H_b$-$T$ phase diagram. The normalization parameters were taken from Table II. The dashed lines are guides to the eye. The arrows point to the location of the tricritical point (TCP) reported in URhGe[46].

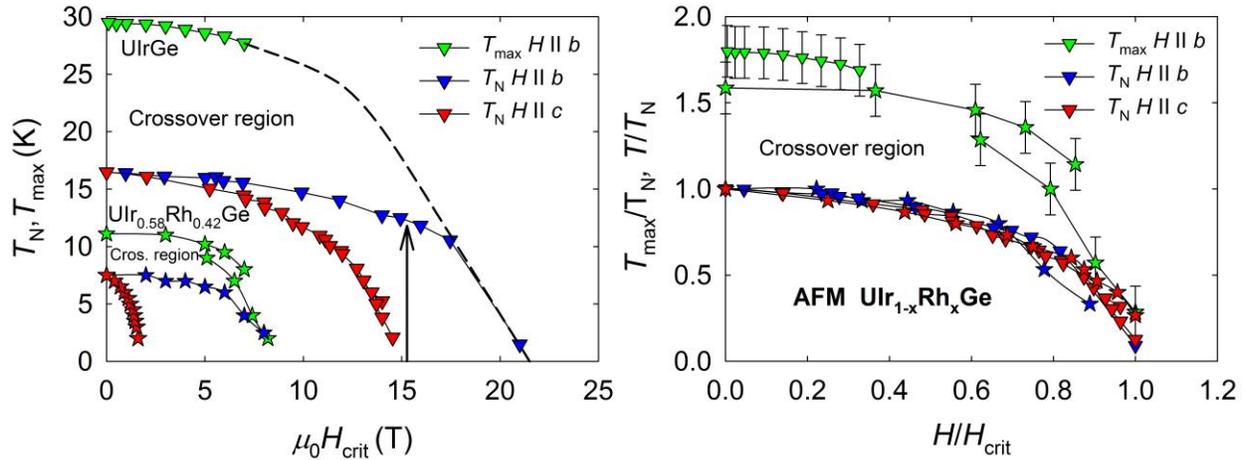

FIG. 17. (Color online) Left) *H-T* phase diagram of UIrGe (triangles) and UIr$_{0.58}$Rh$_{0.42}$Ge (stars) using identical color scheme. Evolution of the $T_N$ of UIrGe was extracted from ref.[14] and $T_{max}$ from our magnetization data. Dashed line tentatively marks evolution of $T_{max}$ in high magnetic fields. Complete dome of the $T_{max}$ of UIr$_{0.58}$Rh$_{0.42}$Ge was constructed using the magnetization and heat capacity data. Righ) Normalized version of the *H-T* phase diagram. The normalization parameters were taken from Table II. The black arrow marks the magnetic field $H_b$ where the peak-like anomaly signaling a first order transition is reported in ref.[14]. The crossover region is clearly demarked by $T_{max}$ and $T_N$ in the magnetic field along the *b* axis.

We constructed an identical set of *H-T* phase diagrams for the AFM part of UIr$_{1-x}$Rh$_x$Ge (Fig. 17) to test the potential propagation of the first order transition near $H_{b,crit}$ from the FM to AFM phase. We omitted UIr$_{0.45}$Rh$_{0.55}$Ge because we cannot exclude the influence of the nascent FM phase detected in the analysis. The *H-T* phase diagram of the AFM part had to be extended by two new parameters $H_{c,crit}$ and especially characteristic temperature $T_{max}$, where $T_{max} > T_N$ in contrast to FM UIr$_{0.43}$Rh$_{0.57}$Ge, UIr$_{0.86}$Rh$_{0.14}$Ge, URhGe, and, we surmise, up to ~UCo$_{0.6}$Rh$_{0.4}$Ge. We performed a similar analysis of the scaling parameters $T_{max}/T_N$, $H_{b,crit}/T_N$ and $H_{b,crit}/T_{max}$ and found overlap between both compounds UIr$_{0.58}$Rh$_{0.42}$Ge and UIrGe (Fig. 17) signaling the presence of an identical order parameter, which is however different from the FM part because of the first order FM/AFM transition. The phase diagram also shows coincidence of the critical field $H_{b,crit}$ for $T_N$ and $T_{max}$ of UIr$_{0.58}$Rh$_{0.42}$Ge. $T_N$ and $T_{max}$ circumscribe a crossover region separating the AFM phase from the PM phase. The $H_{b,crit}$ of UIrGe is too high to see the merger of $T_N$ and $T_{max}$ near $H_{b,crit}$. Taking into account the normalized phase diagram we propose the same scenario here.

### D. Crossover region

The constructed phase diagram in Fig. 17 opens the question as to whether the *b*-axis crossover region circumscribed by $T_N$ and $T_{max}$ is a product of a specific feature of the uranium magnetism or it is related to a heavy fermion phase. It seems that the crossover region is substantially reduced or does not exist in the URhGe because $T_{max} \approx T_C$. Then, critical field $H_{b,crit}$ and $T_{max}$ and $T_C$ are proportional to a constant factor $H_{b,crit}/T_{max} = H_{b,crit}/T_C \approx 1.33$. On the other hand, we surmise a large area of the crossover region existing between the UCoGe - ~UCo$_{0.6}$Rh$_{0.4}$Ge where $T_{max} > T_C$. Very high $H_{b,crit}$ ~ 50 T of UCoGe seems to be connected with $T_{max}$ not having any relation to $T_C$. Identically, $T_{max} > T_N$ in the AFM UIr$_{1-x}$Rh$_x$Ge but both seems to have one common $H_{b,crit}$. Thus, the relation between $T_C$, $T_N$, $T_{max}$ and $H_{b,crit}$ evidently varies through the system and is summarized in Fig. 18. A manuscript supporting the scenario of $T_{max}$ evolution through the UCoGe-URhGe system is in preparation.

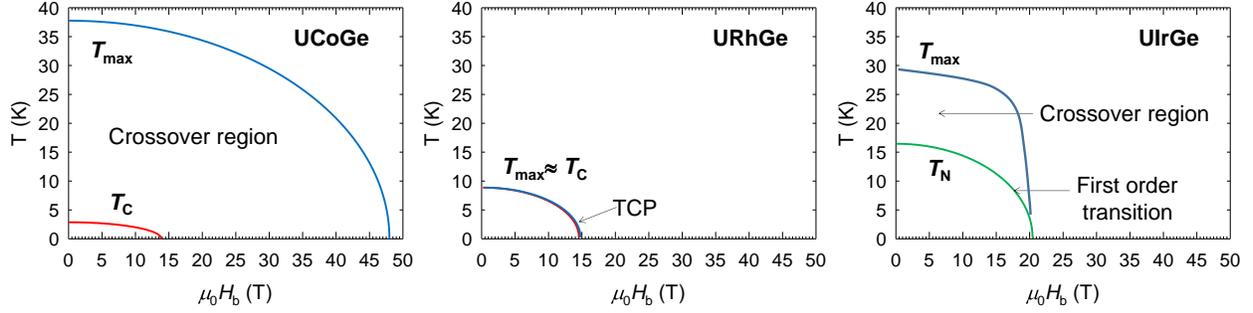

FIG. 18 (Color online) Schematic $H_b$-$T$ phase diagrams of UCoGe, URhGe and UIrGe. The UCoGe diagram was constructed using data from other papers[15, 51], the URhGe diagram is from Ref.[15] and UIrGe diagram is from our data and paper in Ref.[14]. The position of the TCP in URhGe was taken from Ref.[46]. The transformation from the second to first order transition in UIrGe was deduced from the heat capacity data in Ref.[14].

$T_{max}$ is certainly related to the energy scale of AFM ordering with relation $T_{max} \sim 1.8 T_N$ at zero field, as observed in the present paper (Fig. 17). The appearance of $T_{max}$ is a characteristic of an itinerant magnet; a maximum of $\chi$ can be caused due to an increase of spin fluctuations amplitude $\langle \delta m^2 \rangle$ when a certain Fermi surface condition is satisfied[52]. In heavy fermion systems, $T_{max}$ is sometimes connected with the beginning of heavy electron formation. However, in such a case, $T_{max}$ is observed for the easy axis[53-57], which is not the case for the UCoGe-URhGe-UIrGe system. As $T_{max}$ is not observed for the $c$- axis, but exclusively along the $b$-axis, it may be connected rather to a particular shape of the Fermi surface.

It may be interesting to consider UCoGe, which shows high $T_{max} \gg T_C = 2.5$ K. As $T_{max}$ is observed, this compound is considered to be far from the TCP at zero field. If re-entrant superconductivity does not appear at $H_{b,crit}$ in UCoGe, it may be connected to this lack of tricriticality. Actually, in the UCo$_{1-x}$Rh$_x$Ge system, the maximum of $T_C$ appears around UCo$_{0.6}$Rh$_{0.4}$Ge, implying that tricritical and mono- FM fluctuations are enhanced around $x = 1$ and 0, respectively, which reduce $T_C$ but are balanced out around UCo$_{0.6}$Rh$_{0.4}$Ge with the highest $T_C$.

### E. Magnetic fluctuations development in the AFM side of the UIr$_{1-x}$Rh$_x$Ge system

Both AFM compounds UIrGe and UIr$_{0.58}$Rh$_{0.42}$Ge were found with rather low values of the $\gamma$ coefficient compared with common heavy fermion systems. However, the $\gamma$ coefficient is strongly enhanced when the magnetic field is applied along the $b$ and $c$ axes. The $\gamma_c^{2T}{}_{UIr0.58Rh0.42Ge} \approx 125$ mJ/mol K$^2$ (Fig. 9e) is in reasonable agreement with the $\gamma^{band} \approx 110$ mJ/mol K$^2$ (Fig. 13). An enhanced value $\gamma_b^{9T}{}_{UIr0.58Rh0.42Ge} \approx 175$ mJ/mol K$^2$ was found along the $b$-axis exceeding significantly the extrapolated $\gamma^{band}$.

It was shown by Hardy et al.[20] and Miyake et al.[58] that the effective mass $m$ can be described by

$$m^* = m^{band} + m^{**},$$

where $m^{band}$ is the renormalized band mass and $m^{**}$ is the correlated mass associated with the magnetic instability. UIrGe and UIr$_{0.58}$Rh$_{0.42}$Ge are specific cases where extrapolated paramagnetic $\gamma^{band}$ is significantly higher than real $\gamma = 16$ or 70 mJ/mol K$^2$ (Fig. 13). When AFM order vanished in UIr$_{0.58}$Rh$_{0.42}$Ge because of the critical magnetic field $H_{c,crit}$ along the $c$ axis we received a system where $\gamma_c^{2T}{}_{UIr0.58Rh0.42Ge} \approx \gamma^{band}$. On the other hand, when the critical magnetic field $H_{b,crit}$ is applied along the $b$-axis then $\gamma_b^{9T}{}_{UIr0.58Rh0.42Ge} > \gamma^{band}$ giving the magnetic instability term $\gamma^{**} = \gamma_b^{9T}{}_{UIr0.62Rh0.38Ge} - \gamma^{band} \approx 65$ mJ/mol K$^2$. Magnetic field along the $b$-axis enhances fluctuations and a magnetic instability term $\gamma^{**}$ must be taken into account. The origin of the $\gamma^{**}$ term in AFM UIr$_{0.58}$Rh$_{0.42}$Ge may be connected with $T_{max}$ because $T_N$ merges with $T_{max}$ in the vicinity of $H_{b,crit}$ and creates the crossover region (Figs. 17 and 18). We suppose a similar scenario for the parent UIrGe at $H_{b,crit} = 21$ T which must be verified by a high magnetic field experiment. We note that $m^{band}$ in the AFM state may be quite different from that in the PM state, because the Fermi surface in the AFM state is reconstructed due to the AFM magnetic Brillouin zone. The large decrease of the $C_p/T$ in UIrGe below $T_N$ supports this.

## V.  Conclusions

We constructed the magnetic phase diagram of the UIr$_{1-x}$Rh$_x$Ge system and found discontinuity at $x_{crit} = 0.56$ in all the magnetic parameters between the FM and AFM phase typical for the first order transition. QCP is not realized at $x_{crit}$ because of finite $T_C$ and $T_N$. However, magnetic fluctuations are moderately enhanced in the FM limit deduced from the highest $\gamma$ through the FM phase. Magnetic fluctuations are suddenly decreased in the AFM phase. The recovery of the magnetic fluctuations in the AFM compounds is possible in applied magnetic field along the $b$- and $c$- axes. Stronger fluctuations are expected along the $b$ axis probably due to $T_{max}$. We found the dome of the crossover region in the AFM compounds. Based on these findings, we constructed the $d_{U-U}$-$T$ magnetic phase diagram of the UCoGe-URhGe-UIrGe system and schematic phase diagrams of the parent compounds. The relation between $T_{max}$ and $T_C$, $T_N$, and $H_{b,crit}$ seems to be an important feature of the magnetism of the U$T$Ge compounds. Advanced high-magnetic field specific heat and dHvA measurements are desirable to further elucidate the field induced transitions in the UCoGe-URhGe-UIrGe systems.

## Acknowledgments

The authors thank Z. Fisk for fruitful discussion of the experimental data. This paper was supported by Japan Society for the Promotion of Science (JSPS) KAKENHI Grant No. 15H05884 (J-Physics), No. 16K05463, No. JP15H05852, No. 15KK0149, No. 15K05156, and the REIMEI Research Program of Japan Atomic Energy Agency (JAEA). This paper was also supported by a Grant-in-Aid for Scientific Research C (Grant No. 25400386).

# Appendix

Crystal structure parameters of UIrGe.

| $a$ (Å) | $b$ (Å) | $c$ (Å) | $V$ (Å$^3$) | $d_{U-U}$ (Å) |
|---|---|---|---|---|
| 6.8714(4) | 4.3039(3) | 7.5793(5) | 224.15(6) | 3.511 |
| site | $x/a$ | $y/b$ | $z/c$ | Occ. |
| U *(4c)* | 0.0067(20) | 0.25 | 0.7023(19) | 1 |
| Ir *(4c)* | 0.2815(20) | 0.25 | 0.0854(19) | 1 |
| Ge *(4c)* | 0.1831(7) | 0.25 | 0.4132(6) | 1 |

Crystal structure parameters of UIr$_{0.58}$Rh$_{0.42}$Ge.

| $a$ (Å) | $b$ (Å) | $c$ (Å) | $V$ (Å$^3$) | $d_{U-U}$ (Å) |
|---|---|---|---|---|
| 6.8819(5) | 4.3154(3) | 7.5590(6) | 224.494 | 3.511 |
| site | $x/a$ | $y/b$ | $z/c$ | Occ. |
| U *(4c)* | 0.0071(20) | 0.25 | 0.7960(19) | 1 |
| Ir *(4c)* | 0.2823(3) | 0.25 | 0.0863(20) | 0.61(2) |
| Rh *(4c)* | 0.2823(3) | 0.25 | 0.0863(20) | 0.39(2) |
| Ge *(4c)* | 0.1864(6) | 0.25 | 0.4137(6) | 1 |

Crystal structure parameters of UIr$_{0.45}$Rh$_{0.55}$Ge.

| $a$ (Å) | $b$ (Å) | $c$ (Å) | $V$ (Å$^3$) | $d_{U-U}$ (Å) |
|---|---|---|---|---|
| 6.8776(5) | 4.3160(3) | 7.5335(7) | 223.622 | 3.508 |
| site | $x/a$ | $y/b$ | $z/c$ | Occ. |
| U *(4c)* | 0.0067(3) | 0.25 | 0.7959(3) | 1 |
| Ir *(4c)* | 0.2825(5) | 0.25 | 0.0858(4) | 0.43(4) |
| Rh *(4c)* | 0.2825(5) | 0.25 | 0.0858(4) | 0.57(4) |
| Ge *(4c)* | 0.1889(10) | 0.25 | 0.4136(9) | 1 |

Crystal structure parameters of UIr$_{0.43}$Rh$_{0.57}$Ge.

| $a$ (Å) | $b$ (Å) | $c$ (Å) | $V$ (Å$^3$) | $d_{U-U}$ (Å) |
|---|---|---|---|---|
| 6.8747(3) | 4.3141(15) | 7.5476(3) | 223.848 | 3.509 |
| site | $x/a$ | $y/b$ | $z/c$ | Occ. |
| U *(4c)* | 0.0072(9) | 0.25 | 0.7967(8) | 1 |
| Ir *(4c)* | 0.2836(12) | 0.25 | 0.0861(11) | 0.54(1) |
| Rh *(4c)* | 0.2836(12) | 0.25 | 0.0861(11) | 0.47(1) |
| Ge *(4c)* | 0.1875(3) | 0.25 | 0.4141(3) | 1 |

Crystal structure parameters of UIr$_{0.14}$Rh$_{0.86}$Ge.

| $a$ (Å) | $b$ (Å) | $c$ (Å) | $V$ (Å$^3$) | $d_{U-U}$ (Å) |
|---|---|---|---|---|
| 6.8823(5) | 4.3294(3) | 7.5236(6) | 224.1749 | 3.507 |
| site | $x/a$ | $y/b$ | $z/c$ | Occ. |
| U *(4c)* | 0.0077(3) | 0.25 | 0.7950(3) | 1 |
| Ir *(4c)* | 0.2848(6) | 0.25 | 0.0850(5) | 0.15(3) |
| Rh *(4c)* | 0.2848(6) | 0.25 | 0.0850(5) | 0.85(3) |
| Ge *(4c)* | 0.1899(10) | 0.25 | 0.4131(7) | 1 |